\documentclass[letterpaper]{article} % DO NOT CHANGE THIS
\usepackage{aaai25}  % DO NOT CHANGE THIS
\usepackage{times}  % DO NOT CHANGE THIS
\usepackage{helvet}  % DO NOT CHANGE THIS
\usepackage{courier}  % DO NOT CHANGE THIS
\usepackage[hyphens]{url}  % DO NOT CHANGE THIS
\usepackage{graphicx} % DO NOT CHANGE THIS
\usepackage{natbib}  % DO NOT CHANGE THIS AND DO NOT ADD ANY OPTIONS TO IT
\usepackage{caption} % DO NOT CHANGE THIS AND DO NOT ADD ANY OPTIONS TO IT
\usepackage{algorithm}
\usepackage{algorithmic}
\usepackage{subfigure}
\usepackage{mathtools}
\usepackage{multirow}
\usepackage{makecell}
\usepackage{subcaption}
\usepackage{microtype} 
\usepackage{amsmath}
\usepackage{amsthm}
\usepackage{amsfonts} 
\usepackage{booktabs}  
\newtheorem{theorem}{Theorem}
\newtheorem{definition}[theorem]{Definition}
\newtheorem{proposition}[theorem]{Proposition}
\nocopyright
\usepackage{newfloat}
\usepackage{listings}
\DeclareCaptionStyle{ruled}{labelfont=normalfont,labelsep=colon,strut=off} % DO NOT CHANGE THIS
\floatstyle{ruled}
\newfloat{listing}{tb}{lst}{}
\floatname{listing}{Listing}
\title{Rethinking Cancer Gene Identification through \\ Graph Anomaly Analysis}
\author{
    Yilong Zang\textsuperscript{\rm 1}\thanks{Part of this work was carried out while Yilong Zang was affiliated with the German Research Center for Artificial Intelligence (DFKI)}, 
    Lingfei Ren\textsuperscript{\rm 2}\thanks{Corresponding author}, 
    Yue Li\textsuperscript{\rm 3}, 
    Zhikang Wang\textsuperscript{\rm 4}, 
    David Antony Selby\textsuperscript{\rm 5}, 
    Zheng Wang\textsuperscript{\rm 3}, 
    Sebastian Josef Vollmer\textsuperscript{\rm 5}, 
    Hongzhi Yin\textsuperscript{\rm 6}, 
    Jiangning Song\textsuperscript{\rm 4}, 
    Junhang Wu\textsuperscript{\rm 7}
}
\affiliations{
    \textsuperscript{\rm 1}School of Hotel and Tourism Management, The Hong Kong Polytechnic University\\
    \textsuperscript{\rm 2}School of Computing and Artificial Intelligence, Southwestern University of Finance and Economics \\
    \textsuperscript{\rm 3}School of Computer Science, Wuhan University\\
        \textsuperscript{\rm 4}Biomedicine Discovery Institute and Department of Biochemistry and Molecular Biology, Monash University \\
    \textsuperscript{\rm 5}German Research Center for Artificial Intelligence (DFKI) and RPTU Kaiserslautern\\
    \textsuperscript{\rm 6}The University of Queensland \\
     \textsuperscript{\rm 7}College of Information Science and Technology, Shihezi University \\
    yilong.zang@polyu.edu.hk, renlingfei@whu.edu.cn, leeyue@whu.edu.cn, zhikang.wang@monash.edu, selby@informatik.uni-kl.de, wangzwhu@whu.edu.cn, sebastian.vollmer@dfki.de, h.yin1@uq.edu.au, Jiangning.Song@monash.edu, wjh920925@whu.edu.cn
   
}

\begin{document}

\maketitle

\begin{abstract}
Graph neural networks (GNNs) have shown promise in integrating protein--protein interaction (PPI) networks for identifying cancer genes in recent studies. However, due to the insufficient modeling of the biological information in PPI networks, more faithfully depiction of complex protein interaction patterns for cancer genes within the graph structure remains largely unexplored.
This study takes a pioneering step toward bridging biological anomalies in protein interactions caused by cancer genes to statistical graph anomaly. We find a unique graph anomaly exhibited by cancer genes, namely weight heterogeneity, which manifests as significantly higher variance in edge weights of cancer gene nodes within the graph. Additionally, from the spectral perspective, we demonstrate that the weight heterogeneity could lead to the ``flattening out'' of spectral energy, with a concentration towards the extremes of the spectrum. Building on these insights, we propose the HIerarchical-Perspective Graph Neural Network (HIPGNN) that not only determines spectral energy distribution variations on the spectral perspective, but also perceives detailed protein interaction context on the spatial perspective. Extensive experiments are conducted on two reprocessed datasets STRINGdb and CPDB, and the experimental results demonstrate the superiority of HIPGNN. Our code and data are released at \url{https://github.com/zyl199710/HIPGNN}.
\end{abstract}

% Uncomment the following to link to your code, datasets, an extended version or similar.
%
% \begin{links}
%     \link{Code}{https://aaai.org/example/code}
%     \link{Datasets}{https://aaai.org/example/datasets}
%     \link{Extended version}{https://aaai.org/example/extended-version}
% \end{links}

\section{Introduction}

% Cancer genes are broadly defined as genes that confer a selective growth advantage to cells when altered at the genetic, epigenetic, or expression level~\cite{bailey2018comprehensive}.

Identifying cancer genes is a crucial endeavor in both research and clinical practice~\cite{beroukhim2010landscape,martinez2020compendium,bailey2018comprehensive}.
Cancer genes are closely related with protein interactions~\cite{leiserson2015pan}, motivating solutions that integrate the protein--protein interaction (PPI) network for efficient identification~\cite{yang2021efficient,levi2021domino,chitra2022netmix2,yang2023identifying,wang2024dual,wang2023targeting}. Such approaches exploit, for example, multi-omics data and protein interaction information to extract and derive features that distinguish cancer genes.

The aggregation capabilities of graph neural networks (GNNs)~\cite{wu2020comprehensive} have led to notable success in methods for cancer gene identification, based on graph convolutional networks~\cite{schulte2021integration}, Chebyshev graph convolutional works~\cite{peng2022improving} and masked graph autoencoders~\cite{cui2023smg}.
However, these methods only use the PPI network to update the node features by referring to neighbor representations, which do not model the complete biological information within the network.
Therefore there exists a gap: more faithfully depicting complex protein interaction patterns within the graph structure.
%The more vivid depiction of complex protein interaction patterns for cancer genes within the graph structure remains a gap.
% However, these methods superficially use the PPI network to aggregate similar neighboring representations and lack modeling of biological information on the network. The intricate depiction of complex protein interaction patterns for cancer genes within graph structures remains a gap.
% They all propose reasonable solutions to address the scarcity of cancer labels. 
% Unfortunately,\textit{ they do not delve into the statistical differences that cancer genes manifest within the PPI network. In other words, the correlation of cancer genes and protein interactions on graph data has still not been deeply revealed.}

\begin{figure*}[t]
\centering
\includegraphics[width=0.60\linewidth]{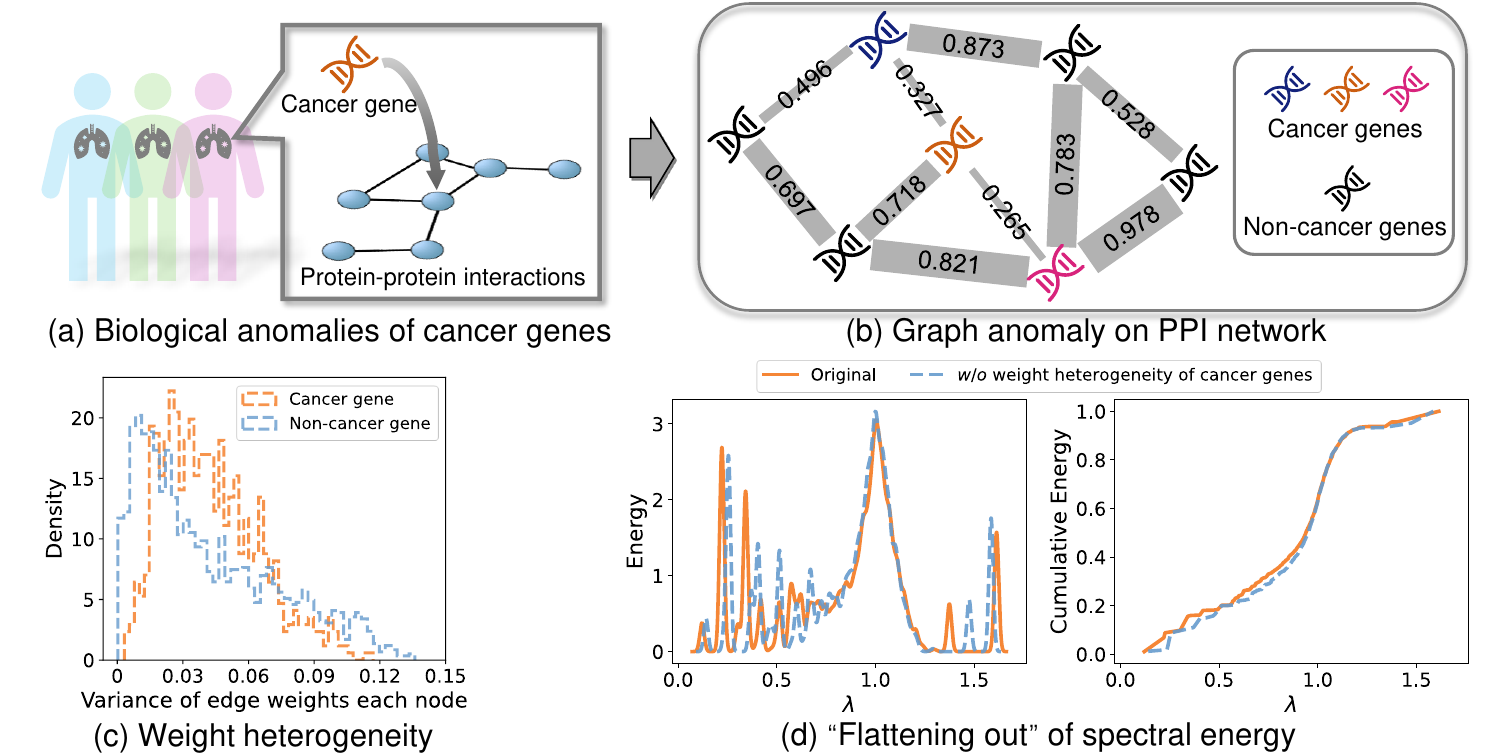}
\vspace{-4pt}
\caption{Overview of our motivation. (a) Cancer genes induce significant biological anomalies in protein interactions. (b) We interpret these anomalies as graph anomaly in the PPI network. (c) Then, we calculate the variance distribution of the edge weights for each gene and investigate the weight heterogeneity of cancer genes from the spatial perspective. (d) Furthermore, we compute and compare the spectral energy distribution with and without weight heterogeneity in cancer nodes and explore the ``flattening out'' of spectral energy from the spectral perspective. We remove the weight heterogeneity of all cancer gene nodes by setting their edge weights to 0.5.}
\label{fig:introduction}
% \vspace{-6pt}
\end{figure*}

Our motivation is as follows: cancer genes induce significant biological anomalies in protein interactions, such as mutations, changes in expression levels, or alterations in protein modifications, as illustrated in Figure~\ref{fig:introduction} (a). 
These biological anomalies can be interpreted as graph anomalies in the PPI network, as shown in Figure~\ref{fig:introduction} (b).
By identifying and analyzing these graph anomalies, we aim to develop a more comprehensive understanding of cancer gene behavior on PPI networks for cancer gene identification. 

% \textcolor{blue}{This and the following graphs have bad coherence. }
Based on this vision, %we perform
our statistical experiments in this paper %and
reveal a distinctive graph anomaly of cancer in the PPI network, which we term \emph{weight heterogeneity}.
As shown in Figure~\ref{fig:introduction} (c), we compute the variance distribution of all edge weights (protein interaction confidence) for each node in a widely used PPI dataset, STRINGdb. It reveals that cancer genes exhibit greater weight variance compared to non-cancer genes. 

Additionally, from the spectral perspective, we demonstrate that weight heterogeneity leads to the ``flattening out'' of the spectral energy, theoretically and experimentally. Figure~\ref{fig:introduction} (d) illustrates the spectral energy distribution with and without weight heterogeneity in cancer nodes using the graph Fourier transform of node attributes. The spectral energy of the original graph (with weight heterogeneity) tends to concentrate more towards the extremes of the spectrum.
We describe this phenomenon as the ``flattening out'' of spectral energy and provide rigorous proof through theoretical analysis. To further illustrate this phenomenon, we also validate it on two synthetic graphs.
Based on the above observations, we recognize that both spatial and spectral perspectives offer information about graph anomaly of cancer. This motivates us to design a cancer gene identification model that integrates both perspectives.

Therefore, we propose an innovative HIerarchical-Perspective Graph Neural Network, termed HIPGNN, to identify cancer genes on the PPI network. HIPGNN can not only discern spectral energy distribution variations to tackle the ``flattening out'' on the spectral perspective, but also perceive detailed protein interaction context for handling weight heterogeneity on the spatial perspective. Specifically, after obtaining the Laplace matrix eigenvalues, we encode the position and proximity of the eigenvalues to integrate the spectral energy distribution information. Following this, we design the proximity-aware spectral graph representation using spectral eigenvalue encoding to update node representations. Finally, we decode the spatial context for node representation by perceiving protein interaction information.

Building on previous works~\cite{schulte2021integration,cui2023smg}, we reprocessed two datasets, STRINGdb and CPDB, which contain real-world PPIs and cancer gene data, to extract more comprehensive protein interaction information. Extensive experiments on these datasets demonstrate the superior performance of the proposed HIPGNN compared to state-of-the-art methods.

\section{Preliminaries}

\begin{figure*}[t]
\centering
\includegraphics[width=0.65\linewidth]{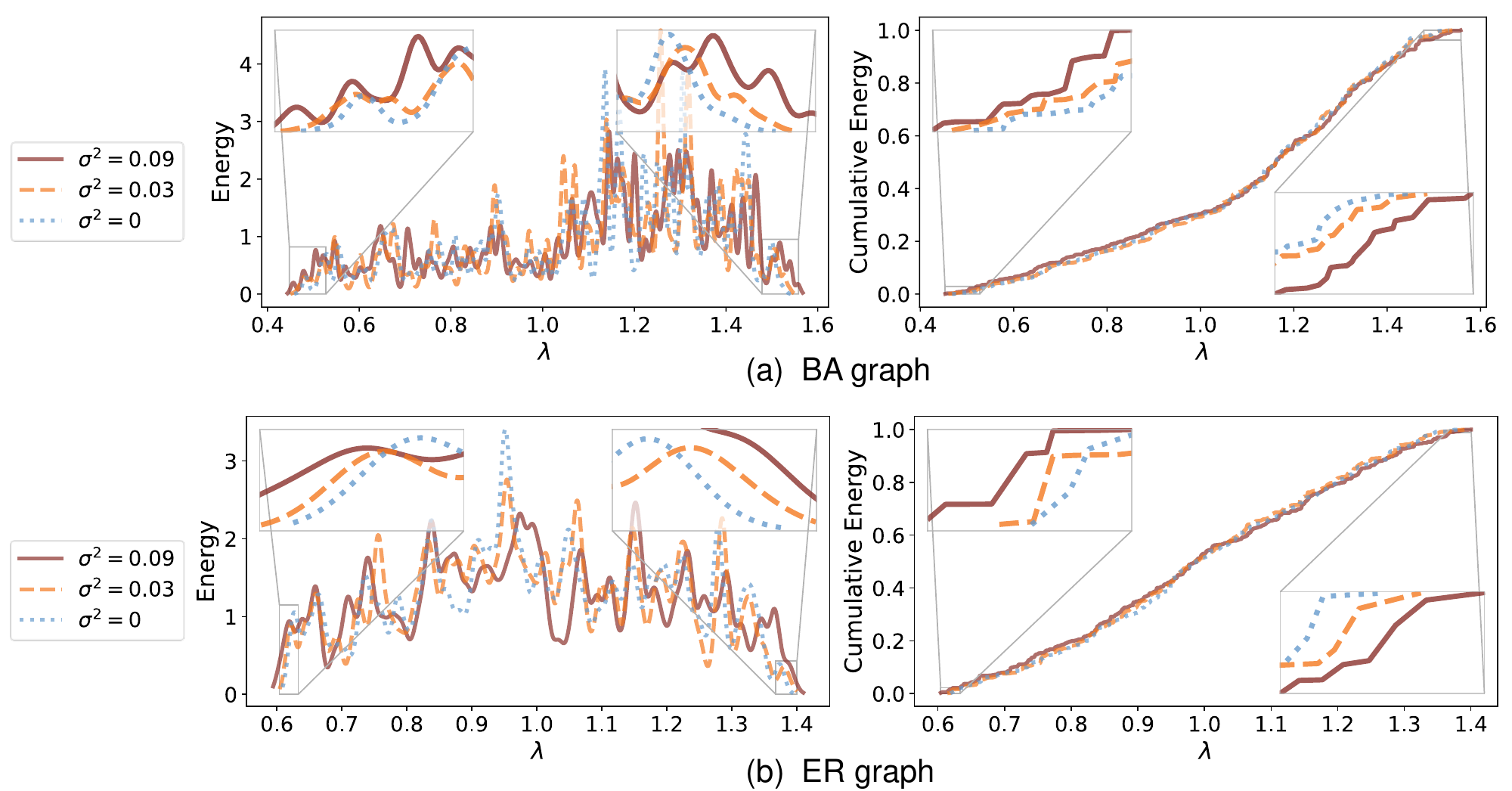}
\vspace{-4pt}
\caption{The distributions of spectral energy and cumulative energy on two synthetic graphs: Barabasi--Albert (BA) graph and Erdős--Rényi (ER) graph. We measure the effect of weight heterogeneity with different weight variances $\sigma^2$. Red solid line means $\sigma^2 = 0.09$, orange dashed line means $\sigma^2 = 0.03$, and blue dotted line means $\sigma^2 = 0$. }
\label{fig:sample}
% \vspace{-6pt}
\end{figure*}

\subsection{Theoretical analysis}

we first provide several necessary definitions and notation.

\paragraph{Weighted graph} We define a weighted graph as $\mathcal{G}_w = \{\mathcal{V},\mathcal{E}, \mathcal{W}, \mathcal{X},  \mathcal{Y} \} $, where $v_i \in \mathcal{V}$ represents the node and $N = \lvert \mathcal{V} \rvert $. 
The node features and labels are denoted as $x_i \in \mathcal{X}$ and $y_i \in \mathcal{Y}$, respectively. The edge $e_{ij} \in \mathcal{E}$ connects nodes $v_i$ and $v_j$, and $w_{ij} \in \mathcal{W}$ is the edge weight of $e_{ij}$. Let $A$ be the corresponding adjacency matrix, where $A_{ij} = w_{ij}$ if there exists a weighted edge. It is worth mentioning that all graphs studied in this paper are undirected graphs, i.e., $A_{ij} = A_{ji}$.

\paragraph{Unweighted graph} An unweighted graph $\mathcal{G}_{uw} = \{\mathcal{V},\mathcal{E}, \mathcal{X}, \mathcal{Y} \}$ is defined similarly to a weighted graph, except that in its adjacency matrix $A$, $A_{ij} = 1$ if there exists an edge.

Given the weight heterogeneity exhibited by cancer genes on the PPI network, we model this phenomenon using a random weighted graph~\cite{khorunzhy2004eigenvalue,ding2010spectral}. Specifically, for an unweighted graph $\mathcal{G}$, we define a set of variables $\{w_{ij}; 1 \leq i < j \leq N\}$ that are independently and identically Gaussian distributed, while assigning the same weight to the symmetric edge weights, making $\mathcal{G}$ a weighted graph. For all $i,j$, $w_{ij} = w_{ji}$, $\mathbb{E}(w_{ij}) = \mu$, and $\mathrm{Var}(w_{ij}) = \sigma^2$. Based on this, we use $\sigma^2$ to measure the degree of weight heterogeneity. Holding $\mu$ constant, we argue that the larger the $\sigma^2$, the higher the weight heterogeneity on the graph.

On the weighted graph $\mathcal{G}$, let $D$ be the diagonal degree matrix. The Laplacian matrix $L$ is defined $L = D - A$ (regular) or $L = I - D^{-1/2}AD^{-1/2}$ (normalized), where $I$ is the identity matrix. $L$ is a symmetric matrix with eigenvalues $0 = \lambda_1 \leq \lambda_2 \leq \cdots \leq \lambda_N$ and a corresponding orthonormal basis of eigenvectors $U = (u_1, u_2, \cdots, u_N)$. Assume that $x = (x_1, x_2, \cdots, x_N)$ is a random signal whose graph Fourier transform is $\hat{x} = U^Tx = (\hat{x}_1, \hat{x}_2, \cdots, \hat{x}_N)$. The spectral energy distribution at $\lambda_k$ is denoted as $f_k(x,L) = \hat{x}^2_k / \sum_{i=1}^N \hat{x}^2_i$. We summarize the following finding from the theory: \textit{\textbf{The weight heterogeneity observed among cancer genes results in ``flattening out'' of spectral energy}, which means that spectral energy is elevated at extremes and lowered in the middle.}

To verify the finding theoretically, we first provide some definitions of the spectral energy distribution:

\begin{definition}
\textnormal{Expectation of spectral energy.} For $\lambda \in {\lambda_1, \lambda_2,\cdots, \lambda_N}$, we define the expectation of the spectral energy on $\lambda$ as:
\begin{equation}
    \mathbb{E}_\lambda(f(x,L)) = \frac{\sum_{k=1}^N\lambda_k\hat{x}^2_k}{\sum_{k=1}^N\hat{x}^2_k}.\nonumber \label{def:ex_e}
\end{equation}
\end{definition}

And Definition~\ref{def:ex_e} can also be converted into the form of Rayleigh quotient~\cite{dong2023rayleigh,gao2023addressing}:
\begin{align}
    \mathbb{E}_\lambda(f(x,L)) &= \frac{\sum_{k=1}^N\lambda_k\hat{x}^2_k}{\sum_{k=1}^N\hat{x}^2_k} = \frac{x^TLx}{x^Tx}   \\
    &= \frac{1}{2} \frac{\sum_{i,j = 1}^N(x_i-x_j)^2w_{ij}}{\sum_{i=1}^N x_i^2}.   \label{eq:ex_e}
\end{align}

Equation~\eqref{eq:ex_e} bridges the energy distribution in the spectral domain with the smoothness of the signal on the graph structure in the spatial domain. It can be seen that if the signal is less smooth, the spectral energy moves to higher points. Further, we define the variance of the spectral energy distribution.

\begin{definition}
\textnormal{Variance of spectral energy.} For $\lambda \in {\lambda_1, \lambda_2,\cdots, \lambda_N}$, The variance of spectral energy on $\lambda$ is defined as:
\begin{equation}
    \mathrm{Var}_\lambda(f(x,L)) = \frac{\sum_{k=1}^N \lambda_k^2 \hat{x}_k^2}{\sum_{k=1}^N \hat{x}_k^2} - \left(\frac{\sum_{k=1}^N \lambda_k \hat{x}_k^2}{\sum_{k=1}^N \hat{x}_k^2}\right)^2.\nonumber
\end{equation}
\end{definition}
A larger variance indicates that the spectral energy disperses more to both sides of the spectrum. Up to this point, we will now state how the weight heterogeneity on the graph structure affects the variance of spectral energy.

\begin{proposition}
Give $L = D - A$ and $\{ w_{ij} \sim \mathcal{N}(\mu, \sigma^2), w_{ij} = w_{ji};  1 \leq i < j \leq N \} $, the expectation of variance of spectral energy with respect to $w$, $\mathbb{E}_w (\mathrm{Var}_\lambda(f(x,L)))$, monotonically increases with the variance of edge weights $\sigma^2$.
\label{pro:ev_e}
\end{proposition} 

The details of the proof process we put in the technical appendix. Proposition~\ref{pro:ev_e} illustrates that a larger variance in the edge weights (weight heterogeneity) on the graph leads to a broader dispersion (``flattening out'') of the spectral energy. Intuitively, disrupting edge weights affects functional connectivity metrics such as effective resistance~\cite{ghosh2008minimizing}, which in turn affects the upper and lower bounds of spectral energy distribution~\cite{barooah2006graph}. 

% Additionally, some important eigenvalues of the Laplacian matrix are closely related to the graph structure. The second smallest eigenvalue $\lambda_2$, associated with Fiedler vectors~\cite{chen2017fiedler}, is a basic component in spectral graph partitioning heuristics and can be maximized by allocating nonnegative edge weights without exceeding a total weight limit~\cite{goring2008embedded}. The largest eigenvalue $\lambda_N$, which may indicate the presence of bottleneck edges, can also be minimized by redistributing the edge weights on the graph~\cite{goring2012graph}.

\subsection{ Validation on synthetic graphs} \label{sec:2.4}

\begin{figure*}[t]
\centering
\includegraphics[width=0.65\linewidth]{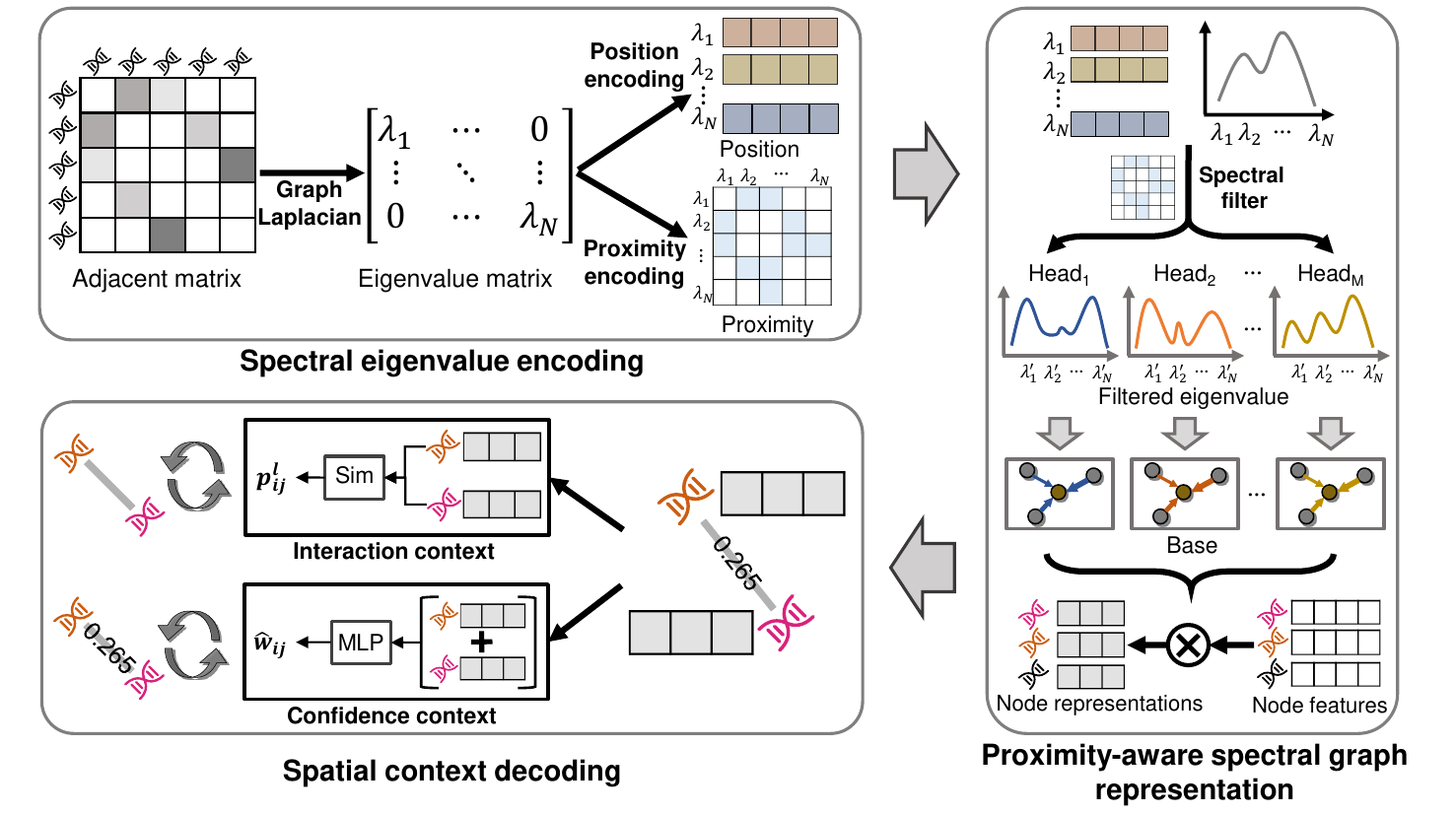}
\vspace{-4pt}
\caption{The overview of the HIPGNN framework. It comprises three modules: the spectral eigenvalue encoding to encode the position and proximity of eigenvalues; the proximity-aware spectral graph representation to fuse eigenvalue position and proximity encoding with spectral filters and get the node representation; the spatial context decoding for perceiving protein interaction information. }
\label{fig:framework}
% \vspace{-6pt}
\end{figure*}

To illustrate our theoretical findings more intuitively, we investigate the ``flattening out'' of spectral energy on Barabasi--Albert (BA)~\cite{albert2002statistical} and Erdős--Rényi (ER)~\cite{erdHos2012spectral} graphs, each with 500 nodes. The BA graph models real-world network properties, while the ER graph has a uniform degree distribution.

To add weight heterogeneity to original graphs, we assign Gaussian independently distributed weights ${w_{ij} \sim \mathcal{N}(\mu, \sigma^2)}$ to all edges, with $w_{ij} = w_{ji}$. The $\sigma^2$ is set to 0.09, 0.03, and 0 (unweighted graph). Given that even with the largest variance, there is only a probability of less than $0.001\%$ for a weight to fall outside the range of $[0, 2]$, we confine the weight values to this interval to ensure realistic weight values. The graph signal is uniformly distributed between 0 and 1.

As shown in Figure~\ref{fig:sample}, we compute and plot the spectral energy distribution (KDE) and the spectral cumulative energy $\eta_k(x, L) = \sum^k_{i=1} \hat{x}^2_i/ \sum^N_{i=1}\hat{x}^2_i$ on the two synthetic graphs. For clarity, we omit the small energy at $\lambda_1 = 0$ and provide a magnified view of the spectrum's extremes. A clear ``flattening out'' of spectral energy is observed on both graphs. We summarize the following observations: (1) For the spectral energy distribution, a larger $\sigma^2$ causes the spectral energy to deviate from the middle and to have lower wave peaks, with a regular arrangement according to $\sigma^2$ size at both ends of the spectrum. (2) For the spectral cumulative energy distribution, an increase in $\sigma^2$ leads to elevated energy in the low-frequency range and reduced energy in the high-frequency range. These trends are particularly pronounced at both ends of the spectrum. Overall, these observations empirically substantiate the Proposition~\ref{pro:ev_e} and provide intuitive insights into the effects of varying $\sigma^2$ on spectral energy distribution.

\subsection{Problem formulation } \label{sec:2.5}

So far, we have formulated the PPI network-based cancer gene identification problem. Existing methods~\cite{schulte2021integration,peng2022improving,cui2023smg} treat protein interactions above a certain confidence threshold as unweighted edges, constructing an unweighted graph. In contrast, we use confidence as edge weights to construct a weighted graph, capturing variations in confidence levels and their correlation with cancer genes in the PPI network.

\paragraph{PPI network based cancer gene identification} This task is regarded as a semi-supervised node classification task. Given a weighted graph $\mathcal{G}_w$ based on a PPI network and some nodes with known labels, our goal is to infer the labels of the remaining nodes, determining whether they are cancer genes.

\section{Method}

Based on the analysis in Preliminaries, cancer genes exhibit a unique graph anomaly, i.e. weight heterogeneity, in the PPI network and show a ``flattening out'' phenomenon in the spectral energy distribution. 
To address the anomaly from both spectral and spatial perspectives simultaneously, we introduce a hierarchical-perspective graph neural network, termed HIPGNN, for cancer gene identification as shown in Figure~\ref{fig:framework}.

% Here, we specifically present the design ideas of the two modules of the model, the proximity-aware spectral graph representation module to tackle the "flattening out" of spectral energy (Section~\ref{sec:3.1}), and the Auxiliary tasks representing and learning to explore the correlation of cancer gene and protein interactions (Section~\ref{sec:3.3}).

% In this section, we introduce our

\subsection{Spectral eigenvalue encoding } \label{sec:3.1}

Most polynomial filter-based spectral GNNs \cite{defferrard2016convolutional,he2022convolutional,he2021bernnet,wang2022powerful} use a fixed polynomial basis for all eigenvalues to approximate arbitrary filters. Nevertheless, these scalar eigenvalue computation methods fall short of expressive capability and cannot capture the ``flattening out'' of the spectral energy well. 
To tackle this issue, we intend to design a more powerful eigenvalue encoding rule to directly reflect the distribution of eigenvalues, such as the spectral gap~\cite{hoffman2021spectral}.

\paragraph{Eigenvalue position encoding} Given a normalized Laplace matrix $L = U\Lambda U^T$ of a weighted graph $\mathcal{G}_w$, we encode each eigenvalue of the matrix $\lambda \in \lambda_1 \leq \lambda_2 \leq \cdots \leq \lambda_N$ from a scalar to a meaningful vector: $\mathbb{R}^1 \rightarrow \mathbb{R}^d$, by using a position encoding function as follows:
\begin{align}
\begin{split}
    \rho_{2i}(\lambda) &= \text{sin}(100\lambda/10000^{2i/d}), \\
    \rho_{2i+1}(\lambda) &= \text{cos}(100\lambda/10000^{2i/d}),
\end{split} 
\end{align}
 where $i$ is an integer and its value domain ranges from $0$ to $d/2-1$. This function forms a multiscale representation of the eigenvalues and has the advantage of filtering arbitrary multivariate continuous functions\cite{bo2022specformer}.
 
 % being able to fit filters with arbitrary multivariate continuous functions\cite{bo2022specformer}. 
 
 \paragraph{Eigenvalue proximity encoding} Furthermore, to intuitively perceive the spectral energy distribution, we propose to encode the proximity between eigenvalues. A proximity matrix is computed by eigenvalue position encodings: $\mathbb{R}^{N\times d} \rightarrow \mathbb{R}^{N \times N}$. Each element of this matrix obtained is as follows:
\begin{align}
    R_{ij} = {\rho(\lambda_i)}^T\rho(\lambda_j),
\end{align}
where there exist the following two theoretical properties of $R_{ij}$ that can be proved.
\begin{proposition} 
    The proximity between $\lambda_i$ to $\lambda_j$, $R_{ij}$, is determined by $\lambda_i - \lambda_j$. \label{pro:position1}
\end{proposition}
\begin{proposition} 
    The $R_{ij}$ is undirected. \label{pro:position2}
\end{proposition}
The two propositions (proof in the technical appendix) illustrate that proximity matrix $R_{ij}$ can effectively capture and represent the spectral energy distribution variations, which further enables the GNN to process the spectral energy ``flattening out''.
% provides the GNN with useful information for processing the spectral energy "flattening out".

\subsection{ Proximity-aware spectral graph representation \label{sec:3.2}}
After generating the valuable eigenvalue encoding, we utilize it in the spectral graph representation. Then we propose a Transformer-based\cite{vaswani2017attention} proximity-aware spectral graph representation to fuse eigenvalue position and proximity encoding with spectral filters and get the node representation.

\paragraph{Proximity-aware spectral filter} Unlike using a regular transformer to design spectral filters~\cite{bo2022specformer}, we introduce the eigenvalue proximity information to the attention computation process for designing trainable spectral filters. Given the initial representation which concatenates eigenvalues with their encodings:$Z = \begin{bmatrix}
\lambda_1 \lVert p(\lambda_1) , \cdots ,\lambda_n \lVert p(\lambda_n)
\end{bmatrix}^T \in \mathbb{R}^{N \times (d+1)}$, an innovative attention computation function is proposed as follows:
\begin{align}
\begin{split}
        Q &= ZW_m^Q, \quad K = ZW_m^K,\quad V = ZW_m^V,  \\
     Z'_m &= \text{Attention}(Q,K,V)  = \text{Softmax}(\frac{QK^T+R}{\sqrt{d_q}})V,  
\end{split}
\end{align}
where $d_q$ is the dimension of each head, and $m$ represent the $m$-th head. We include the proximity matrix $R$ as part of attention for learning the global distribution of eigenvalues. Afterward, the representation $Z'_m$  of each head is used as a spectral filter to compute new eigenvalues as $\lambda_m' = \phi(Z'_m W_\lambda)$, where $\lambda_m' \in \mathbb{R}^{N \times 1} $ is the $m$-th eigenvalue vector after the spectral filtering.

\paragraph{Learnable bases} After obtaining $M$ vectors of filtered eigenvalues, we use a feed-forward network (FFN) in the standard Transformer layer to create the learnable bases for graph convolution. The reconstruction and concatenating processes can be formulated as follows:
% Reconstructing new bases and concatenating them is defined as follows:
\begin{gather}
    S_m = U\text{diag}(\lambda_m')U^T, \quad  \hat{S} = \text{FFN}([I_N\lVert S_1 \lVert \cdots \lVert S_M]),
\end{gather}
where $I_N \in \mathbb{R}^{N \times N}$ denotes the unit matrix.

\paragraph{Graph convolution} Eventually, we regard each dimension of the node features as a graph signal and multiply it with the combined Laplace matrix base $\hat{S}$: 
\begin{gather}
    \hat{X}_{:,i}^{l-1} = \hat{S}_{:,:,i}X^{l-1}_{:,i}, \quad X^{l} = \sigma(\hat{X}^{l-1}W_x^{l-1}) + X^{l-1},
\end{gather}
where $\sigma()$ is activation and $X^{l}$ is the node representation in the $l$-th layer.

\subsection{Spatial context decoding \label{sec:3.3}} 

Recalling our findings, in addition to the ``flattening out'' of the spectral energy, we also observe weight heterogeneity within the weighted graph. This indicates that the information about the protein interaction context over the spatial domain is also helpful in distinguishing such anomalies. Motivated by this hypothesis, we decode the protein interaction context to correlate different nodes on graph data. Therefore, after obtaining the node representations, we design the spatial context decoding module to perceive the protein interaction and confidence information in the spatial domain. 

% Specifically, we focus on perceiving the context of nodes from both protein interactions and interaction confidence to capture and learn the association among node representations on graph data,

\paragraph{Interaction context perception} Given node representations $X^l$, we compute the interaction probability between $x^l_i$ and $x^l_j$ by cosine similarity and then leverage cross entropy to approximate the interactions on the graph:
\begin{align}
\begin{split}
            p_{ij}^l &= \text{cos}(x_i^l, x_j^l), \\
        \mathcal{L}_l &= \sum_{\mathclap{(i,j)\in \hat{\mathcal{E}}}} (y^l_{ij}\text{log}(p_{ij}^l) +  (1-y^l_{ij})\text{log}(1-p_{ij}^l)), 
\end{split}
\end{align}
where the set $\hat{\mathcal{E}}$ contains the edges $\mathcal{E}$ on the graph and the negatively sampling edges from the original dataset. if $(i, j)$ is negatively sampling edge, $y_{ij}^l = 0$.

\paragraph{Confidence context perception} More importantly, the model needs to perceive protein interaction confidence to tackle weight heterogeneity. Given the node representations $X^w$, the MLP model and MSE are utilized to predict confidence scores between $x^w_i$ and $x^w_j$ as well as to compute losses, respectively:
\begin{align}
\begin{split}
            \hat{w}_{ij} &= \text{MLP}((x^w_i + x^w_j)/2), \\ 
        \mathcal{L}_w &= \sum_{\mathclap{(i,j)\in \hat{\mathcal{E}}_\text{train}}} \text{MSE}(\hat{w}_{ij}, w_{ij}).
\end{split}
\end{align}
In $(i,j) \in \hat{\mathcal{E}}$, if $(i, j)$ is negatively sampling edge, $w_{ij} = 0$.

It is worth mentioning that we set up node representing channels independent of cancer gene identification for the above two perception modules. We use multiple standard transformer models to obtain separate node representations for each channel: $X^n$, $X^l$, and $X^w$.

\paragraph{Cancer gene identification} Here, we proceed with the loss function for cancer gene identification.
% calculate the loss of our task. 
We feed $X^n$ to the MLP with sigmoid function to get the cancer gene node probability $p^n$. The weighted cross-entropy loss is used to alleviate the challenge from label imbalance as follows:
% labeled imbalances: 
\begin{align}
        \mathcal{L}_n = \sum_{i\in \mathcal{V}_{train}} (\gamma y_i \text{log}{p_i^n} +(1-y_i)\text{log}(1-p_i^n)),
\end{align}
where $\mathcal{V}_{train}$ is the training set of nodes $\mathcal{V}$, and $\gamma$ is the ratio of cancer gene nodes ($y_i = 1$) to non-cancer gene nodes ($y_i = 0$) in the training set. At last, we sum all the losses with weights to get the total loss:$\mathcal{L} = \alpha\mathcal{L}_n + \beta \mathcal{L}_l + \gamma \mathcal{L}_w$. 

\begin{table*}[t]
\small
\centering
\setlength\tabcolsep{1.3pt} 
\renewcommand{\arraystretch}{0.550}
\resizebox{0.72\linewidth}{!}{
 \begin{tabular}{c c p{0.9cm}<{\centering}p{0.9cm}<{\centering}p{1.0cm}<{\centering} p{0.9cm}<{\centering}p{0.9cm}<{\centering}p{1.0cm}<{\centering} p{0.9cm}<{\centering}p{0.9cm}<{\centering}p{1.0cm}<{\centering} p{0.9cm}<{\centering}p{0.9cm}<{\centering}p{0.9cm}<{\centering}}
\toprule
\multicolumn{1}{c}{\multirow{4}*{Graph}}& \multicolumn{1}{c}{\multirow{4}*{Method}}& \multicolumn{6}{c}{STRINGdb} & \multicolumn{6}{c}{CPDB} \\
 \cmidrule(lr){3-8} \cmidrule(lr){9-14}
\multicolumn{1}{c}{}&\multicolumn{1}{c}{}& \multicolumn{3}{c}{20\%} & \multicolumn{3}{c}{80\%} & \multicolumn{3}{c}{20\%} & \multicolumn{3}{c}{80\%} \\
 \cmidrule(lr){3-5}\cmidrule(lr){6-8}\cmidrule(lr){9-11}\cmidrule(lr){12-14}
 &  & \small{AUC} & \small{F1} & \small{AP} & \small{AUC} & \small{F1} & \small{AP} & \small{AUC} & \small{F1} & \small{AP} & \small{AUC} & \small{F1} & \small{AP} \\
\midrule
\multicolumn{1}{c}{\multirow{9}*{Unweighted}} & GCN
 & 81.68 & 72.09 & 64.16 
 & 87.99 & 77.43 & 75.61 
 & 81.84 & 73.54 & 67.55 
 & 82.48 & 73.67 & 69.16 \\
\multicolumn{1}{c}{} & GAT
 & 81.67 & 71.62 & 58.88 
 & 85.25 & 73.35 & 68.52 
 & 80.16 & 70.66 & 61.02 
 & 84.50 & 76.26 & 68.70 \\
\multicolumn{1}{c}{} & GraphSAGE
 & 84.37 & 73.75 & 65.48 
 & 87.09 & 78.98 & 72.13 
 & 78.02 & 69.32 & 62.99 
 & 85.96 & 78.36 & 76.31 \\
\multicolumn{1}{c}{} & Chebnet
 & 83.35 & 73.20 & 66.47 
 & 86.44 & 78.33 & 73.39 
 & 75.70 & 66.77 & 59.29 
 & 82.00 & 71.72 & 68.32 \\
 \cmidrule(lr){2-14}
\multicolumn{1}{c}{} & EMOGI
 & 79.06 & 64.27 & 59.74 
 & 86.88 & 70.18 & 73.22 
 & 80.84 & 67.10 & 66.94 
 & 80.84 & 68.25 & 64.00 \\
\multicolumn{1}{c}{} & MTGCN
 & 84.30 & 73.97 & 66.82 
 & 86.90 & 76.05 & 73.89 
 & 77.25 & 69.62 & 60.95 
 & 83.71 & 68.84 & 70.22 \\
\multicolumn{1}{c}{} & SMG
 & \textbf{89.81} & 78.62 & 75.69 
 & 90.80 & 79.74 & 77.43 
 & 84.75 & 72.34 & 70.83 
 & 86.57 & 77.77 & 77.83 \\
\multicolumn{1}{c}{} & HIPGNN
 & 89.08 & 78.17 & 75.56 
 & 90.81 & 81.71 & 79.66 
 & \textbf{87.99} & 79.13 & 78.07
 & 87.80 & 79.87 & 77.40 \\
\midrule
\multicolumn{1}{c}{\multirow{5}*{Weighted}} & GCN
 & 81.65 & 72.46 & 63.91 
 & 87.17 & 76.25 & 74.22 
 & 81.79 & 73.91 & 67.04 
 & 82.97 & 73.70 & 68.93 \\
\multicolumn{1}{c}{} & GAT
 & 77.99 & 69.93 & 54.49
 & 85.85 & 74.50 & 72.67 
 & 74.48 & 42.19 & 57.71 
 & 85.51 & 75.69 & 68.55 \\
\multicolumn{1}{c}{} & Chebnet
 & 83.64 & 73.91 & 66.58 
 & 87.17 & 78.03 & 74.72 
 & 76.00 & 66.32 & 59.81   
 & 83.35 & 74.34 & 69.82 \\
 \cmidrule(lr){2-14}
\multicolumn{1}{c}{} & HIPGNN 
 & 88.39 & \textbf{79.60} & \textbf{76.13} 
 & \textbf{91.18} & \textbf{83.33} & \textbf{81.21} 
 & 87.88 & \textbf{79.38} & \textbf{78.13} 
 & \textbf{89.66} & \textbf{80.91} & \textbf{79.71} \\
\bottomrule
\end{tabular}
}
\caption{Performance on the two datasets under different percentages of the training data. (\%)}
\label{tab:performance}
\end{table*}

\subsection{Complexity analysis} 

Considering the large size of the graph, we intend to use only a few important eigenvalues as inputs to the model in order to greatly reduce the computational complexity. By analyzing the spectral energy distribution, we believe that the eigenvalues at both extremes are more effective in encoding the ``flattening out'' of the spectral energy.

Therefore, we introduce a hyperparameter $q$ to adjust for only considering the first $q$ small and the last $q$ large eigenvalues. Ultimately, the complexity of HIPGNN is $\mathcal{O}(2q^2d_1 + 2q^2M + Nd_1L + Nd_1^2 + N^2d_2 + 2E^2d_2 + Nd_2^2 + 2Ed_2^2)
$, where $N$ and $E$ are the nodes and edges of the weighted graph, $M$ and $L$ denote the number of filters and layers, and $d_1$ and $d_2$ represent the hidden dimensions of graph layer and node representation.

\section{Experiments}

\subsection{Experimental setup }

\paragraph{Datasets} Based on previous works~\cite{schulte2021integration,cui2023smg}, we extract richer protein interaction information on two widely used PPI datasets with confidence~\cite{szklarczyk2021string,kamburov2009consensuspathdb}, and integrate cancer gene data to construct two datasets. We name these two datasets directly after the PPI databases: STRINGdb and CPDB. Unlike previous works that used fixed threshold confidence to construct unweighted graphs, HIPGNN directly leverages protein confidence as edge weights to construct weighted graph.

\paragraph{Metrics} We choose AUC, F1 (macro), and AP for model performance evaluation. AUC measures the area under the ROC curve, providing a global assessment across all classification thresholds. F1 (macro) is the unweighted average of F1 scores for both categories, suitable for imbalanced datasets. AP is the area under the precision-recall curve, and is considered the most important metric for cancer gene identification~\cite{schulte2021integration}.

\paragraph{Baselines} 
The baseline methods can be categorized into two groups: firstly, general GNN-based models including GCN~\cite{kipf2016semi}, GAT~\cite{velivckovic2018graph}, GraphSAGE~\cite{hamilton2017inductive}, and Chebnet~\cite{defferrard2016convolutional}; and secondly, state-of-the-art cancer gene identification methods including EMOGI~\cite{schulte2021integration}, MTGCN~\cite{peng2022improving} and SMG~\cite{cui2023smg}. We also implement GCN, GAT, and Chebnet on weighted graph and HIPGNN on unweighted graph.

\subsection{Performance comparison}

Table~\ref{tab:performance} presents the results of HIPGNN and other baseline methods with training ratios of 20\% and 80\%. From the table, we draw the following conclusions.

\paragraph{Importance of spectral graph representation} Only Chebnet and HIPGNN outperform on weighted graphs compared to unweighted ones, highlighting that edge weights can negatively impact models like GCN, which function as low-pass filters. This demonstrates the effectiveness of appropriate spectral filters in addressing weight heterogeneity.

\paragraph{Importance of spatial context} HIPGNN shows a significant performance boost at a 20\% training ratio, particularly on the CPDB dataset, outperforming SMG by 7.30\% in AP. This indicates that spatial context in protein interactions aids in identifying unknown cancer genes, especially when labels are sparse.

\paragraph{Superiority of HIPGNN} HIPGNN consistently outperforms other models across most metrics, effectively handling weight heterogeneity to distinguish cancer genes. Notably, HIPGNN improves AP by 0.44\% on STRINGdb and 7.30\% on CPDB at a 20\% training ratio, and by 3.78\% on STRINGdb and 1.88\% on CPDB at an 80\% training ratio, compared to SMG.

Due to space constraints, subsequent experiments focus on the STRINGdb dataset, with CPDB results provided in the technical appendix.

\subsection{Ablation analysis }

\begin{figure}[t]
\centering
\includegraphics[width=0.92\linewidth]{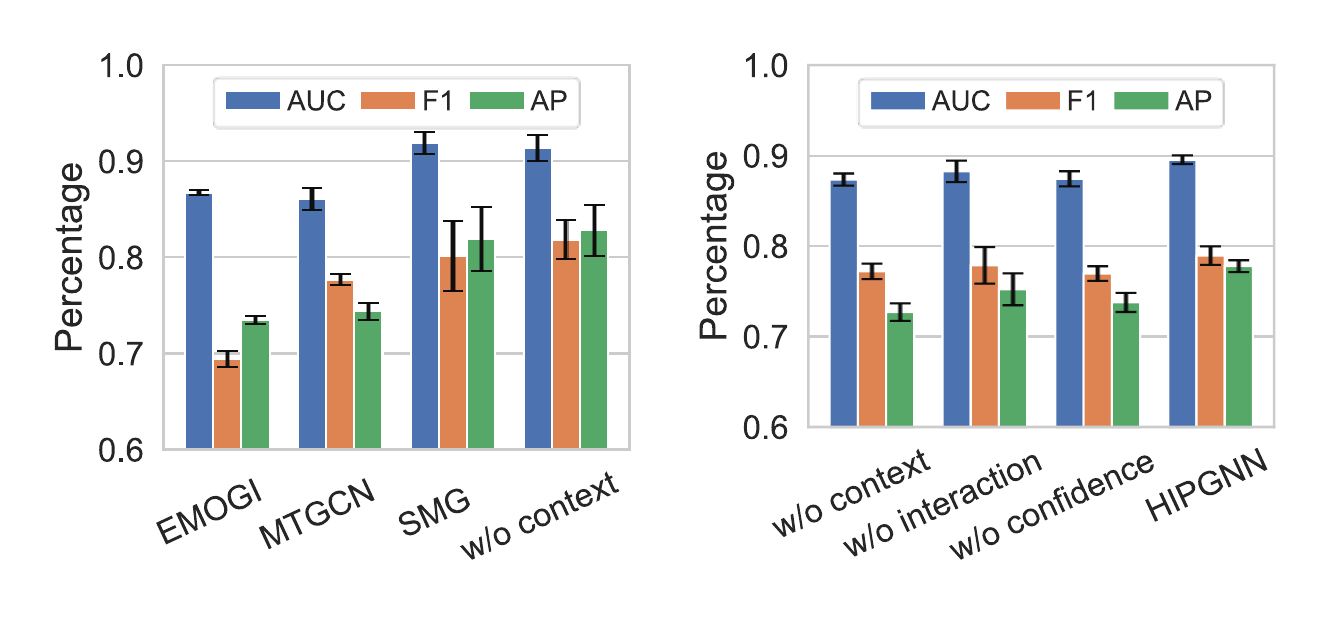}
    % \vspace{-6pt}
    \caption{Ablation analysis of spectral graph representation (left) and spatial context decoding (right) on the STRINGdb dataset.}
    \label{fig:ablation}
 \vspace{-4pt}
\end{figure}

\begin{figure}[t]
\centering
\includegraphics[width=0.86\linewidth]{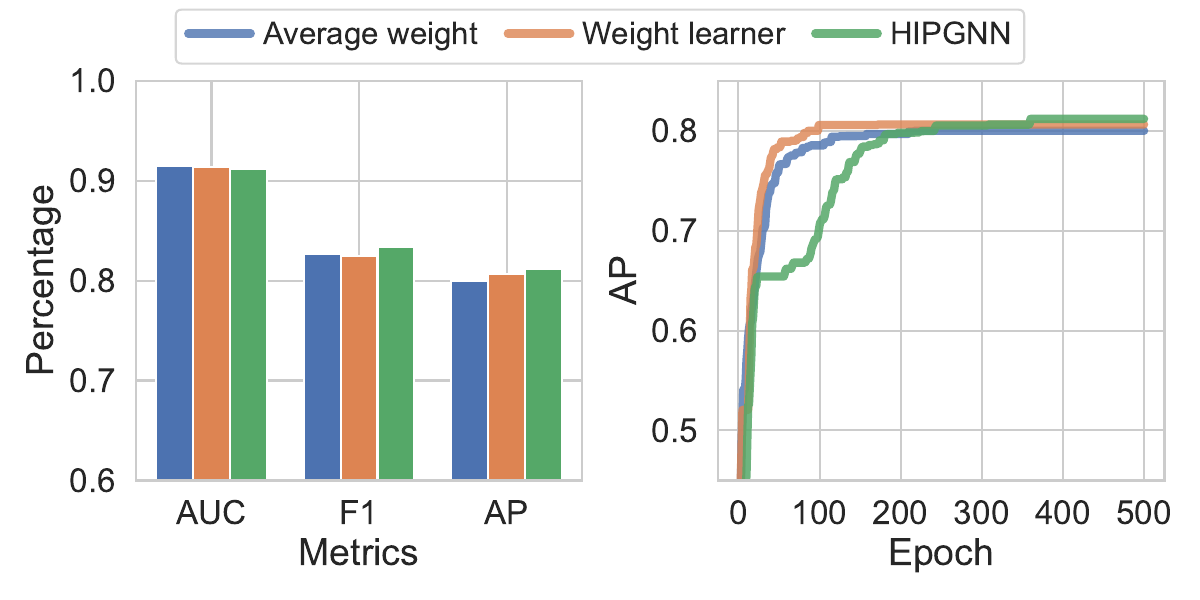}
\caption{Comparison of the three model metrics as well as the variation of the best AP metric in the test set under three different loss weight schemes.}
\label{fig:loss_stringdb}
\vspace{-4pt}
\end{figure}

\begin{figure}[t]
\centering
\includegraphics[width=0.92\linewidth]{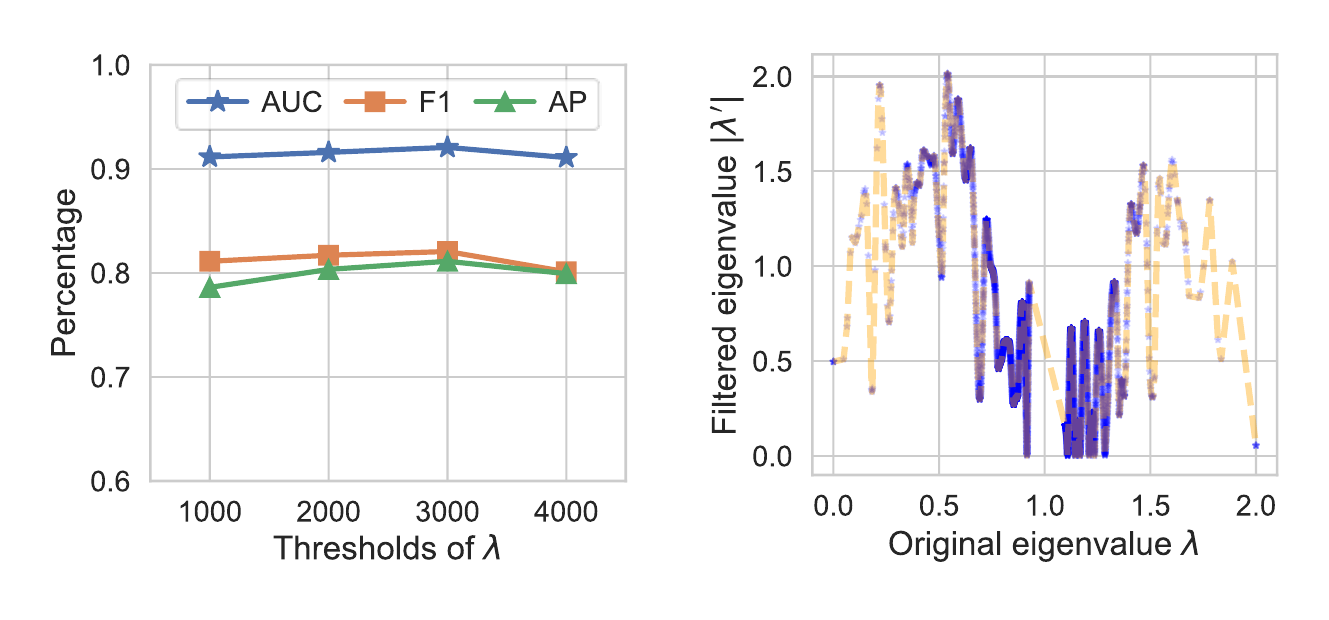}
    % \vspace{-6pt}
    \caption{The eigenvalue thresholds analysis (left) and the spectral filter visualization (right).}
    \label{fig:lambda&filter}
 \vspace{-4pt}
\end{figure}

\paragraph{Proximity-aware spectral graph representation} We examine the impact of the spectral graph representation module with spectral eigenvalue encoding in HIPGNN. We compare EGOGI, MTGCN, SMG, and HIPGNN (without spatial context decoding) using five-fold cross-validation at an 80\% training ratio. The left subfigure of Figure~\ref{fig:ablation} shows box plots of the results, where HIPGNN with only spectral graph representation still outperforms, highlighting the effectiveness of spectral eigenvalue encoding.

\paragraph{Spatial context decoding} For decoding protein interaction contexts, we consider both interaction and confidence contexts for node representations. We evaluate the contribution of these contexts to HIPGNN using four variants: (1) Without context: removes both interaction and confidence contexts; (2) Without interaction: removes interaction context; (3) Without confidence: removes confidence context; (4) HIPGNN: the original model. The right subfigure of Figure~\ref{fig:ablation}, using a 20\% training ratio and five-fold cross-validation, shows that both contexts improve HIPGNN’s performance, with confidence context being particularly impactful.

\subsection{Parameter analysis}

\paragraph{Loss weights} For the final loss computation, we used $\alpha$, $\beta$, and $\gamma$ to weight the protein interaction context, interaction confidence context, and cancer gene label loss, respectively. We empirically set $\alpha = 0.01$ on STRINGdb and $\alpha = 0.02$ on CPDB, with $\beta = 2/3(1- \alpha)$ and $\gamma = 1/3(1- \alpha)$. To validate this, we compared two other schemes: Average weight (uniformly setting all weights to $1/3$) and Weight learner, which uses a Bayesian learnable loss function~\cite{li2020multi,peng2021novel}:
$\mathcal{L} = \frac{1}{\alpha_l^2}\mathcal{L}_n + \frac{1}{\beta_l^2}\mathcal{L}_l + \frac{1}{\gamma_l^2}\mathcal{L}_w + 2\text{log}(\alpha_l\beta_l\gamma_l), $
where $\alpha_l$, $\beta_l$, and $\gamma_l$ are learnable parameters. Figure~\ref{fig:loss_stringdb} shows the three model performance metrics and the best AP metric variation over 500 epochs under the three schemes. Empirical weights achieved the best results. Additionally, setting $\alpha$ smaller aids in the convergence of cancer gene labeling loss. We observed that Average weight and Weight learner fall into local optima early, while our scheme continues optimizing, with metrics possibly improving beyond 500 epochs.

\paragraph{Thresholds of eigenvalue} We introduced the parameter $q$ to control the selection of the first $q$-small and last $q$-large eigenvalues input into the model, thereby regulating its complexity. We evaluated HIPGNN's performance with $q$ values ranging from 1,000 to 4,000. As shown in the left subfigure of Figure~\ref{fig:lambda&filter}, setting $q$ to 3,000 yielded the best performance. Increasing $q$ beyond this point resulted in decreased performance, aligning with our initial observation that the spectral energy tends to "flatten out" more at the spectrum's ends.

\subsection{Visualization of spectral filter }

We applied a spectral filter to obtain the filtered eigenvalues in the proximity-aware spectral graph representation. In the right subfigure of Figure~\ref{fig:lambda&filter}, we visualize the relationship between the filtered and original eigenvalues. The blue stars depict the distribution of eigenvalues, connected by a yellow dashed line. The figure shows that the spectral filter prioritizes eigenvalues near the spectrum's ends over those in the middle. This suggests that spectral eigenvalue coding effectively addresses the key phenomenon in PPI networks caused by cancer genes: the "flattening out" of spectral energy.

\section{Conclusion}

This work takes a pioneering step toward bridging significant biological anomalies in protein interactions caused by cancer genes to the statistical graph anomaly. We identify a unique graph anomaly in cancer genes, termed weight heterogeneity, which leads to the “flattening out” of spectral energy. In response, we propose a novel model, HIPGNN, for the identification of cancer genes.

\paragraph{Broader impact.} This work has the potential to benefit both the bioinformatics and network science fields. It not only lays a new theoretical foundation for cancer gene identification but also offers a fresh perspective and direction for research in graph anomaly detection.

\paragraph{Limitations.} The phenomenon of weight heterogeneity was observed only in cancer genes on two PPI networks. Further validation on other PPI networks is necessary to refine this observation. Additionally, exploring other real-world scenarios where weight heterogeneity occurs could provide more validation datasets for graph anomaly detection.

\bibliography{aaai25}

\appendix

\section{A. Proof of Equation 1} \label{app:eq1}

\begin{proof}
Given the laplace matrix $L = D - A $ and the eigenvalue matrix $\Lambda$ of $L$, we have:
\begin{align}
\mathbb{E}_\lambda(f(x,L)) &= \frac{\sum_{k=1}^N\lambda_k\hat{x}^2_k}{\sum_{k=1}^N\hat{x}^2_k} =  \frac{\hat{x}^T\Lambda\hat{x}}{\hat{x}^T\hat{x}}   \nonumber \\
 &= \frac{\hat{x}^T U^T L U \hat{x}}{\hat{x}^T U^T U\hat{x}} =\frac{x^TLx}{x^Tx}.  \nonumber
 % &= \frac{x^TL}{}\frac{\sum_{(i,j)\in \mathcal{E}}(x_i-x_j)^2w_{ij}}{\sum_{i=1}^N x_i^2}. \nonumber
\end{align}
For $x^TLx$, we have:
\begin{align}
  x^TLx &= x^TDx - x^TAx = \sum^N_{i = 1} d_ix_i^2 - \sum^N_{i,j = 1} x_ix_jw_{ij} \nonumber \\
  &= \frac{1}{2}(\sum^N_{i= 1}d_ix_i^2 - 2\sum^N_{i,j = 1}x_ix_jw_{ij} + \sum^N_{j = 1}d_jx_j^2) \nonumber \\
  &= \frac{1}{2}\sum_{i,j =1}^N (x_i - x_j)^2 w_{ij}. \nonumber 
\end{align}
Hence, we get:
\begin{align}
  \frac{x^TLx}{x^Tx} &= \frac{1}{2} \frac{\sum_{i,j =1}^N (x_i - x_j)^2 w_{ij}}{\sum_{i=1}^N x_i^2}. \nonumber 
\end{align}
\end{proof}

\section{B. Proof of Proposition 3} \label{app:ev_e}

\begin{proof}
For $\mathrm{Var}_\lambda(f(x,L))$, similar to the reasoning in Equation 1, we have:
\begin{align}
    \mathrm{Var}_\lambda(f(x,L)) &= \frac{\sum_{k=1}^N \lambda_k^2 \hat{x}_k^2}{\sum_{k=1}^N \hat{x}_k^2} - \left(\frac{\sum_{k=1}^N \lambda_k \hat{x}_k^2}{\sum_{k=1}^N \hat{x}_k^2}\right)^2 \nonumber \\
    &= \frac{x^TL^2x}{x^Tx} - {\left( \frac{x^TLx}{x^Tx} \right)}^2. \nonumber
\end{align}
We first discuss $x^TL^2x$:
\begin{align}
    x^TL^2x &= x^T (D-A)^2 x = x^T (D^2 - 2DA + A^2) x  \nonumber \\
    &= \sum_{i=1}^N x_i^2 \left( \sum_{j=1}^N w_{ij} \right)^2 \nonumber \\  
    &- 2 \sum_{i,j=1}^N x_i x_j \left( \sum_{k=1}^N w_{ik} \right)w_{ij}  \nonumber\\
    &+ \sum_{i,j=1}^N x_i x_j \left( \sum_{k=1}^N w_{ik} w_{kj} \right).  \nonumber
\end{align}

Then we discuss the relationship between $\mathbb{E}_w(x^TL^2x)$ and $\sigma^2$. Since, $w_ij$ obeys independent Gaussian distributions and $\mathbb{E}(w_{ij}^2) = \text{Var}(w_{ij}) + \mathbb{E}(w_{ij})^2 = \sigma^2 + \mu^2$, then only $\mathbb{E}(w_{ij}^2)$ is positively related to $\sigma^2$. Then we simplify  the three parts of the last equation above to the $\sigma^2$-dependent terms:
\begin{align}
 \sum_{i=1}^N x_i^2 \left( \sum_{j=1}^N w_{ij}  \right)^2  &\xrightarrow{\sigma^2}  \sum_{i,j=1}^N x_i^2w_{ij}^2,  \nonumber \\
 2 \sum_{i,j=1}^N x_i x_j \left( \sum_{k=1}^N w_{ik} \right)w_{ij} &\xrightarrow{\sigma^2} 2\sum_{i,j=1}^N x_ix_jw_{ij}^2, \nonumber \\
 \sum_{i,j=1}^N x_i x_j \left( \sum_{k=1}^N w_{ik} w_{kj} \right) &\xrightarrow{\sigma^2}
 \sum_{i=1}^N x_i^2 \left( \sum_{k=1}^N w_{ik}w_{ki} \right) \nonumber \\
 &= \sum_{i,j=1}^N x_i^2w_{ij}^2. \nonumber
\end{align}

Here we abuse the $\propto$ as a positive correlation. Then we consider $w_{ij} = w_{ji}$ and bring them back to $\mathbb{E}_w(x^TL^2x)$:
\begin{align}
x^TL^2x &\xrightarrow{\sigma^2} \sum_{\mathclap{1\leq i < j \leq N}} (x_i^2+ x_j^2)w_{ij}^2  -
2\sum_{\mathclap{1\leq i < j \leq N}} (x_ix_j + x_jx_i)w_{ij}^2 \nonumber \\ 
&+ \sum_{\mathclap{1\leq i < j \leq N}} (x_i^2 +x_j^2) w_{ij}^2 \nonumber \\
&= 2 \sum_{\mathclap{1\leq i < j \leq N}} (x_i - x_j)^2 w_{ij}^2, \nonumber \\
\mathbb{E}_w(x^TL^2x) &\propto 2 \sum_{\mathclap{1\leq i < j \leq N}} (x_i - x_j)^2 \sigma^2.          \nonumber
\end{align}   

Similarly, we simplify $(x^TLx)^2$ and bring it into $\mathbb{E}_w\left( (x^TLx)^2 \right)$.
\begin{align}
&(x^TLx)^2 = \frac{1}{4} \left( \sum_{i,j = 1}^N (x_i - x_j)^2 w_{ij} \right)^2  \nonumber \\
&\xrightarrow{\sigma^2}\!\frac{1}{4}\!\left( \sum_{i,j = 1}^N (x_i\!-\!x_j)^4 w_{ij}^2\!+\!\sum_{i,j = 1}^N (x_i\!-\!x_j)^2(x_j\!-\!x_i)^2 w_{ij}w_{ji} \right) \nonumber \\
&= \sum_{\mathclap{1\leq i < j \leq N}} (x_i - x_j)^4 w_{ij}^2, \nonumber 
\end{align}
\begin{align}
\mathbb{E}_w \left( (x^TLx)^2 \right) \propto \sum_{\mathclap{1\leq i < j \leq N}} (x_i - x_j)^4 \sigma^2.          \nonumber
\end{align}

Finally, we calculate $\mathbb{E}_w (\mathrm{Var}_\lambda(f(x,L)))$:
\begin{align}
&\mathbb{E}_w (\mathrm{Var}_\lambda(f(x,L))) \nonumber \\ &= \mathbb{E}_w \left( \frac{x^TL^2x}{x^Tx} \right) -  \mathbb{E}_w \left( \frac{(x^TLx)^2}{(x^Tx)^2}  \right) \nonumber \\
&\propto\!\frac{\sigma^2}{(x^Tx)^2}\!\left(  \enspace 2  \sum_{\mathclap{1\leq i < j \leq N}} (x_i\!-\!x_j)^2\!\left( \sum_{k=1}^N x_k^2 \right)\!-\!\sum_{\mathclap{1\leq i < j \leq N}} (x_i\!-\!x_j)^4  \right)   \nonumber \\
&=   \frac{\sigma^2}{(x^Tx)^2} \enspace \sum_{\mathclap{1\leq i < j \leq N}}  (x_i - x_j)^2   \left( 2\sum_{k=1}^N x_k^2 - (x_i - x_j)^2 \right) \nonumber \\
&=  \frac{\sigma^2}{(x^Tx)^2} \enspace \sum_{\mathclap{1\leq i < j \leq N}}  (x_i\!-\!x_j)^2 \left( \sum_{k=1}^N x_k^2\!+\!\sum_{\mathclap{k \neq i , k \neq j}} x_k^2 \!+\!(x_i\!+\!x_j)^2    \right).   \nonumber
\end{align}
It is intuitively clear that all terms except $\sigma^2$ are positive, which means that $\mathbb{E}_w (\mathrm{Var}_\lambda(f(x,L)))$ is monotonically increasing for $\sigma^2$.
\end{proof}

\section{C. Proof of Proposition 4 and Proposition 5 } \label{app:position}

\begin{proof}
For simplicity, we let $\omega_k = 100/10000^{2k/d} $, then we get:
\begin{align}
    R_{ij} &= {\rho(\lambda_i)}^T\rho(\lambda_j) \nonumber \\ &= \sum_{k=0}^{d/2-1} (\text{sin}(\omega_k \lambda_i)\text{sin}(\omega_k \lambda_j)  + \text{cos}(\omega_k \lambda_i)\text{cos}(\omega_k \lambda_j)    ) \nonumber \\
    &= \sum_{k=0}^{d/2-1} \text{cos}(\omega_i(\lambda_i - \lambda_j)). \nonumber
\end{align}
Similarly, we can prove $R_{ji} = \sum_{k=0}^{d/2-1} \text{cos}(\omega_k(\lambda_j - \lambda_i)) = R_{ij}$, and therefore prove the two propositions.
\end{proof}

\section{D. Details of experimental setup} \label{app:setup}

\subsection{Datasets} \label{app:dataset}

\begin{table*}[ht]
\footnotesize
\centering
\caption{Statistical details of two datasets, STRINGdb and CPDB.}
\label{tab:dataset}
 \setlength\tabcolsep{5pt} 
 \renewcommand{\arraystretch}{0.7}
\begin{tabular}{c cc cc c}
\toprule
Dataset & Nodes&  Cancer genes   &  Non-cancer genes & Edges & Confidence range \\
\cmidrule(lr){1-1}\cmidrule(lr){2-4}\cmidrule(lr){5-6}
STRINGdb&  13,179 &  783 &  2,415& 908,908& $[0.6, 1]$\\
Preliminary&  2,928 &  752 & 2,176 & 22,395 & $[0, 1]$\\
\midrule
CPDB&  13,627 &  796 &  2,187& 578,140& $[0.4, 1]$\\
Preliminary& 2,637  &  775 & 1,862& 28,085& $[0, 1]$\\
\bottomrule
\end{tabular}
\vspace{-4pt}
\end{table*}

For cancer gene attributes, we gathered mutation, copy number, DNA methylation, and gene expression data for 29,446 samples across 16 cancer types from the TCGA database~\cite{weinstein2013cancer}. In line with the EMOGI~\cite{schulte2021integration}, we analyzed these cancer types using DNA methylation data and preprocessed batch effect-corrected gene expression data from both tumor and normal tissues to derive nodal features and labels for the multi-omics data.

For PPI networks, we collected two public databases: STRINGdb~\cite{szklarczyk2021string} and CPDB~\cite{kamburov2009consensuspathdb}. Unlike previous approaches, we employed a lower confidence threshold to capture protein interactions, treating them as a weighted graph. Specifically, in the STRINGdb database, we considered all protein interactions with a confidence level above 0.6, while in the CPDB database, the threshold was set at 0.4. For the negative sampling process, even though we set a threshold to select the interaction confidence, we still excluded all confidence interactions and randomly sampled a number of negative edges equal to the actual edges on the datasets. Such an operation makes our negatively sampled edges more closely resemble the real sense of the no-interaction relation.
After aligning the node attributes and labels as well as the PPI network data, we obtain the final two datasets STRINGdb and CPDB, containing complete weighted graph information. 

For the dataset in the preliminaries, we extracted all data with labels of whether they are cancer genes or not, considering protein interactions between them at all confidence levels. The statistical details of all datasets are shown in Table~\ref{tab:dataset}.

\subsection{Baselines} 
We use two groups of baselines to compare with HIPGNN. The first group is normal GNN models including:
\begin{itemize}
    \item GCN~\cite{kipf2016semi} is the typical graph neural network method that aggregates features from its direct neighbors and itself.
    \item GAT~\cite{velivckovic2018graph} uses the attention coefficient to aggregate node features.
    \item GraphSAGE~\cite{hamilton2017inductive} is an inductive GNN method, that generates node embeddings by sampling and aggregating features from a fixed-size neighborhood of each node.
    \item Chebnet~\cite{defferrard2016convolutional} is a classic spectral GNN method that employs Chebyshev polynomials to define convolutional filters on graphs.
\end{itemize}

The second group is state-of-the-art cancer gene identification methods including:
\begin{itemize}
    \item EMOGI~\cite{schulte2021integration} is a GCN-based cancer gene identification method that employs a loss function with weights to alleviate label imbalance.
    \item MTGCN~\cite{peng2022improving} is a Chebnet-based GNN method that utilizes multi-task learning to enhance cancer gene identification efficiency.
    \item SMG~\cite{cui2023smg} leverages the self-supervised masked graph autoencoder model to tackle the limited cancer gene labels.
\end{itemize}
We also improve GCN, GAT, and Chebnet applied on the weighted graph and apply our HIPGNN on the unweighted graph.

\subsection{Implementation Details} 

All experiments were performed on the L4 GPU resource in the Google colaboratory~\cite{bisong2019google}. Except as mentioned in the original paper, all the learning rate is set to 0.01, the hidden layer for the graph learning process is set to 128, the hidden layer for node representation is set to 64, and the number of layers in a GNN is set to $2$. For the loss weights, we empirically set $\alpha = 0.01$ on the STRINGdb dataset, and $\alpha = 0.02$ on the CPDB dataset. Then we compute $\beta = 2/3(1-\alpha)$ and $\gamma = 1/3(1-\alpha)$. The training ratio is set to 20\% and 80\% to model performance more adequately. To uniformly evaluate the performance of models, we show the best performance of the AP metric throughout 500 training epochs in the test set. And the random seed is fixed as 42.

\section{E. Supplementary experimental results on CPDB dataset}

\begin{figure*}[t]
\centering
\includegraphics[width=1\linewidth]{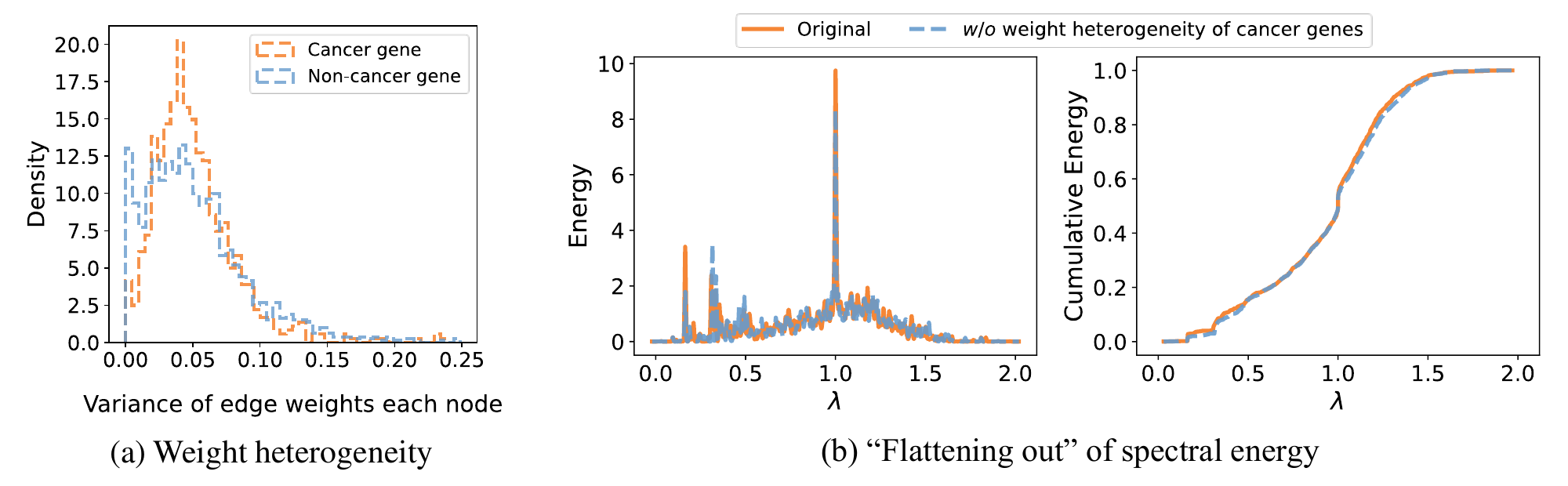}
\caption{The preliminary experiments of weight heterogeneity (a) and ``flattening out'' of spectral energy (b) on the CPDB dataset. }
\label{fig:CPDB}
\end{figure*}

On the CPDB dataset, We validate the weight heterogeneity and ``flattening out'' of spectral energy as shown in Figure~\ref{fig:CPDB}. We can also observe the phenomenon of weight heterogeneity. The ``flattening out'' of the spectral energy is more pronounced in the low-frequency region than in the high-frequency region. This may be due to the fact that the edge weights of real-world data are not exactly ideally Gaussian randomly distributed, making some experimental results not obvious. 

\begin{figure*}[t]
\centering
\includegraphics[width=0.7\linewidth]{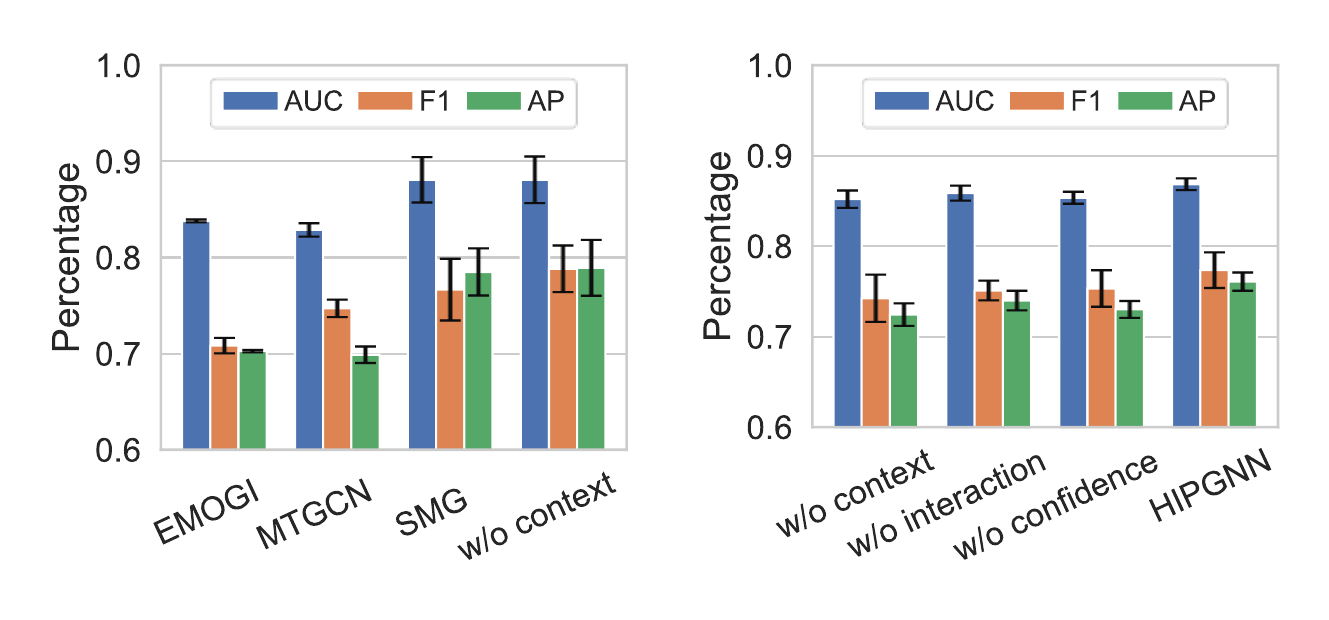}
\caption{Ablation analysis of proximity-aware spectral graph representation (left) and spatial context decoding (right) on CPDB dataset.}
\label{fig:ab_cpdb}
\end{figure*}

\begin{figure*}[t]
\centering
\includegraphics[width=0.7\linewidth]{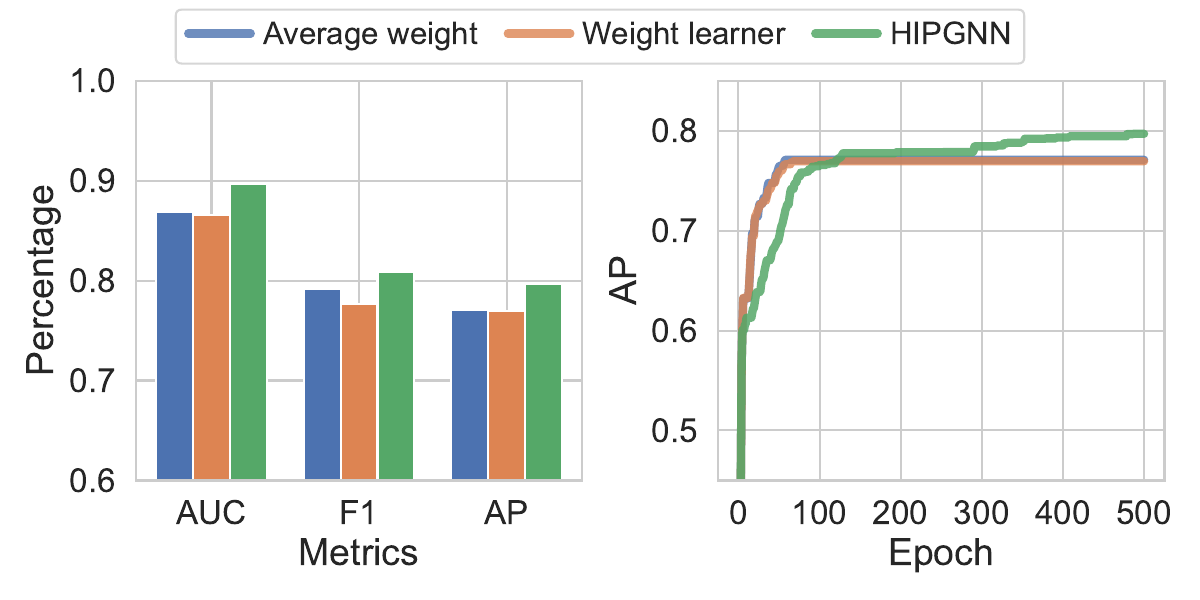}
\caption{Comparison of the three model metrics as well as the variation of the best AP metric in the test set under three different loss weight schemes on the CPDB dataset.}
\label{fig:loss_cpdb}
\end{figure*}

\begin{figure*}[t]
\centering
\includegraphics[width=0.7\linewidth]{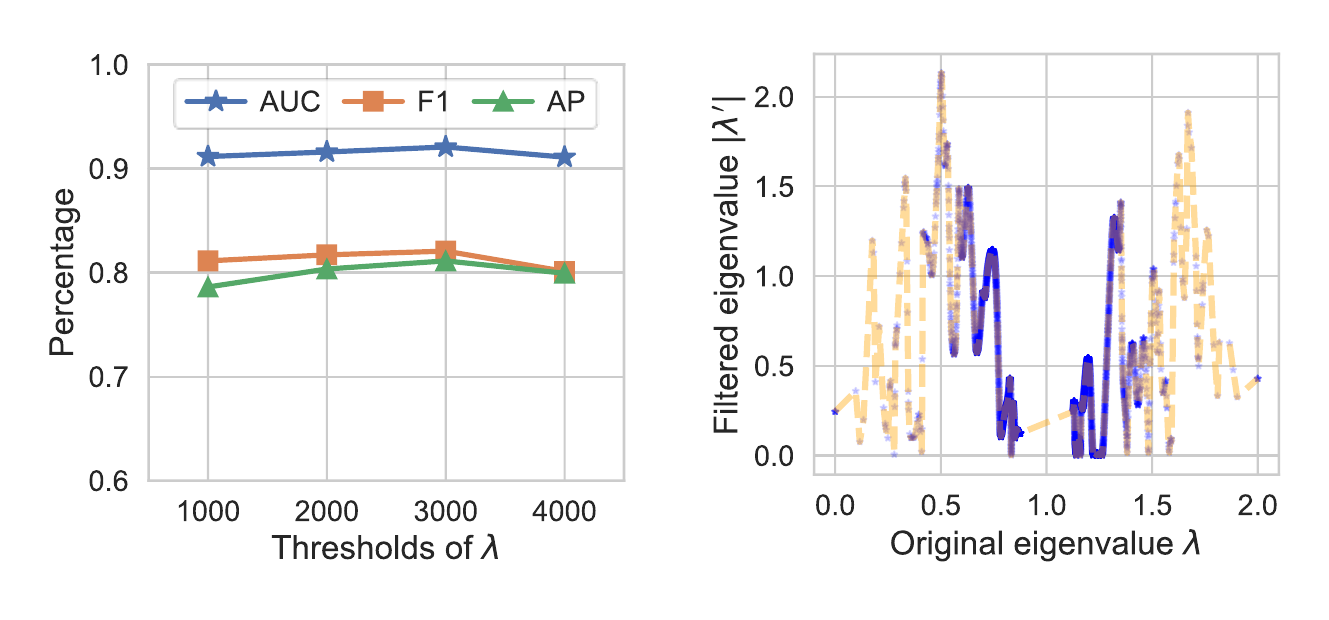}
\caption{The eigenvalue thresholds analysis (left) and the spectral filter visualization (right) on the CPDB dataset.}
\label{fig:lambda_cpdb}
\end{figure*}

In addition, as shown in Figures~\ref{fig:ab_cpdb},~\ref{fig:loss_cpdb}, and~\ref{fig:lambda_cpdb}, we demonstrate the ablation analysis, parametric analysis, and visualization of spectral filter experiments on the CPDB dataset. We find consistent phenomena and conclusions all over the CPDB dataset.

\section{F. Biomedical discussion}

\subsection{Explanation of weight heterogeneity}

We attribute the weight heterogeneity of cancer genes on PPI networks to tumor heterogeneity. Tumor heterogeneity stems from genetic and environmental differences that affect the cellular phenotype~\cite{marusyk2010tumor,kar2009human}. This phenotype is determined by intracellular proteins that regulate messaging~\cite{pietzner2021mapping}. Oncogenes in tumors interact with a variety of other proteins through complex cellular functions and pathways~\cite{phan2021atm}, forming a more intricate network of protein interactions~\cite{hanahan2011hallmarks}. In contrast, non-cancer genes have relatively stable expression processes, and their interactions with other proteins are more consistent and similar.

\begin{figure*}[t]
\centering
\includegraphics[width=1.0\linewidth]{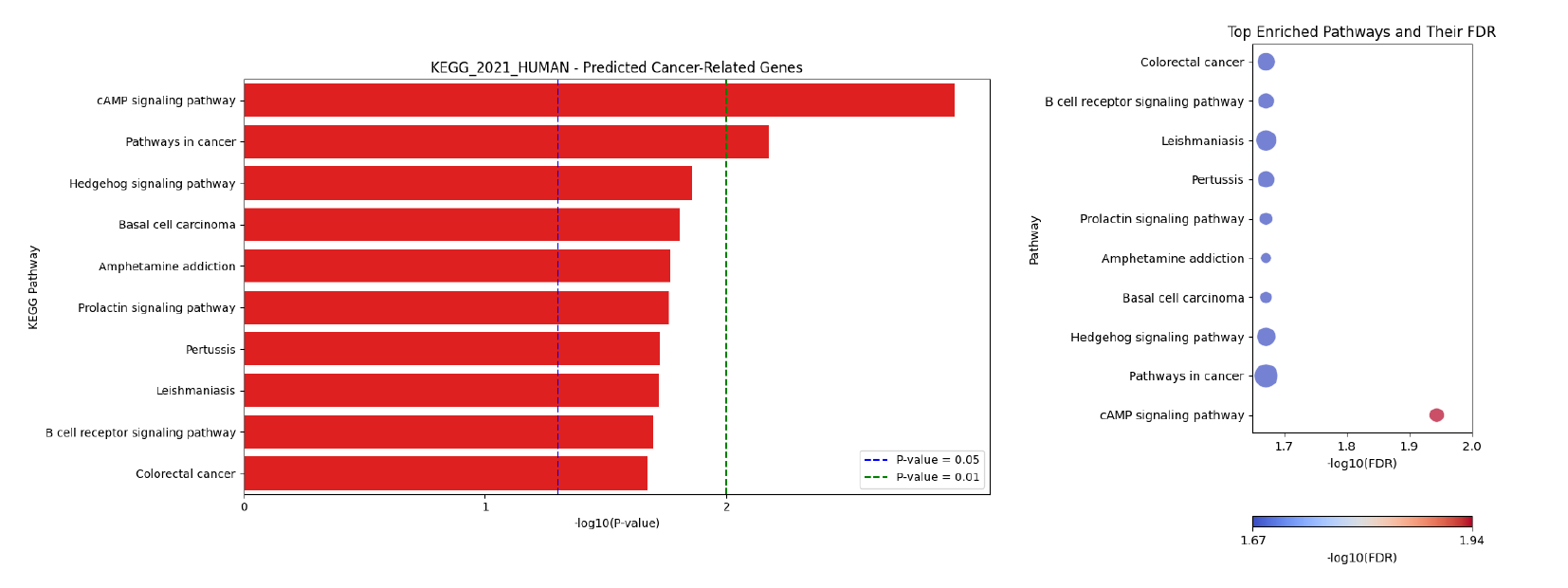}
\caption{Visualization of gene set enrichment analysis using the top-ranked genes from the attribution results within the KEGG pathways. Pathways are ranked by the P-value (left). Pathways are ranked by the FDR (right).}
\label{fig:bio}
\end{figure*}

\subsection{Case study}
We conducted pathway enrichment analysis, in an attempt to better understand the specific roles of top-ranked predicted gene set, providing important clues for further research and treatment. To achieve this, we utilized the Enrichr API~\cite{kuleshov2016enrichr} with the KEGG-2021-human dataset~\cite{kanehisa2021kegg}, conducting a hypergeometric test for pathway enrichment analysis. The left subfigure of Figure~\ref{fig:bio} displays that the enrichment results of the top-ranked gene sets were predominantly influenced by cancer-related pathways. 

The right subfigure of Figure~\ref{fig:bio} shows that the False Discovery Rate (FDR) values range from approximately 0.011 to 0.021, indicating a low false discovery rate and high statistical significance of the identified results.

\section{G. Related work} \label{app:relatedwork}

We present related work in two directions: PPI network based cancer gene identification and graph anomaly detection.

\subsection{PPI network based cancer gene identification}

The goal of this task is to identify cancer genes using cancer gene-related attributes and PPI networks. EMOGI~\cite{schulte2021integration} pioneered the use of Graph Convolutional Network (GCN) integrating multi-omics data as well as PPI networks to identify cancer genes. Then MTGCN~\cite{peng2022improving} considered the Chebyshev graph convolutional and utilized multi-task learning to enhance cancer gene identification efficiency. SMG~\cite{cui2023smg} utilized a masked graph autoencoder to separately learn graph structure information and node classification to address the challenge of limited cancer gene labels. 

The differences between HIPGNN and existing methods can be summarized as follows: (1) HIPGNN considers richer protein interaction information, i.e., protein interaction confidence as the weight of the edges to construct a weighted graph; whereas all existing methods perform graph representation only on unweighted graphs. (2) HIPGNN analyzes the essential properties of cancer genes on PPI networks from the biological information perspective, and then designs an explainable and efficient GNN model; this is not available in existing methods.

\subsection{Graph anomaly detection}

The graph anomaly is defined as an abnormal or unusual pattern of nodes, edges, or subgraphs in the graph data~\cite{akoglu2015graph}. From the perspective of the task, current graph anomaly detection methods can be divided into two main categories, node-level anomaly detection, and graph-level anomaly detection.

\paragraph{Node-level anomaly detection} This task focuses on identifying the distribution of anomalies exhibited by the anomalous nodes. One of the most important anomalies is node heterogeneity~\cite{zheng2022graph}, which means that nodes with different labels tend to link. FAGNN~\cite{bo2021beyond} employs a self-gating mechanism for the adaptive fusion of different signals in the message passing process. H2GCN~\cite{zhu2020beyond} identifies a set of key designs, i.e. ego- and neighbor-embedding separation, higher-order neighborhoods, and combination of intermediate representations, for GNN. ACM~\cite{luan2022revisiting} studies heterogeneity in terms of node similarity after aggregation and adaptively exploits aggregation, diversity and identity channels in each GNN layer. GHRN\cite{gao2023addressing} demonstrated that heterogeneity is positively related to frequency, and proposed a graph heterophily resistant network, which is equipped with a label-aware high-frequency indicator. In addition to the study of node heterogeneity, AO-GNN~\cite{huang2022auc} proposes to decouple the AUC maximization process on GNN into a classifier parameter searching and an edge pruning policy searching process to solve the label-imbalance as well as the node heterogeneity. BWGNN~\cite{tang2022rethinking} finds that node attribute anomalies lead to a shift in spectral energy towards higher frequencies and proposes the beta wavelet GNN.

\paragraph{Graph-level anomaly detection} The task focuses on analyzing and identifying the anomalies of different subgraphs. 
OCGIN~\cite{zhao2021using} first explores graph-level anomaly detection and analyzes the performance flip-flop of several methods on graph-categorized datasets. iGAD~\cite{zhang2022dual} shows that anomalous substructures lead to graph anomalies. It proposes an anomalous substructure-aware deep random walk kernel and a node-aware kernel to capture topology and node features. RQGNN~\cite{dong2023rayleigh} investigates the spectral differences between anomalous and non-anomalous graphs and proposes a Rayleigh quotient GNN to explore the inherent spectral features of anomalous graphs.

In summary, the current discussion of graph anomalies includes node attribute anomaly~\cite{tang2022rethinking}, structural anomaly~\cite{zheng2022graph,bo2021beyond,zhu2020beyond,luan2022revisiting,gao2023addressing,zang2023don}, subgraph anomaly~\cite{zhang2022dual,dong2023rayleigh}, and label anomaly~\cite{huang2022auc}. To the best of our knowledge, this study is the first to address edge weight anomalies in weighted graphs. This unique graph anomaly, termed weight heterogeneity, is characterized by higher variance in the edge weights of anomalous nodes.

Furthermore, leveraging spectral GNNs~\cite{tang2022rethinking,gao2023addressing,dong2023rayleigh} is currently a prominent trend in graph anomaly detection. In this study, we innovatively discuss weight heterogeneity in the spectral domain, proving that it leads to the ``flattening out'' of spectral energy. Based on this insight, we propose an advanced hierarchical-perspective GNN from both spectral and spatial perspectives, achieving state-of-the-art performance in cancer gene identification.

\section{H. Contributions} 
Our contributions can be described in terms of both cancer gene identification and graph anomaly detection:
\begin{itemize}
\item[-] \textbf{For cancer gene identification.} We provide an in-depth analysis of the properties of cancer genes on PPI networks and propose weight heterogeneity to outline this property. In addition, we propose a hierarchical-perspective GNN, HIPGNN, which achieves state-of-the-art performance on two reprocessed publicly available cancer gene identification datasets.

\item[-] \textbf{For graph anomaly detection.} We propose a unique graph anomaly, weight heterogeneity, described as an elevated variance of edge weights of anomalous nodes. Through experimental and theoretical validation, we find that weight heterogeneity leads to ``flattening out'' of the spectral energy. Furthermore, for this anomaly, we propose a state-of-the-art method HIPGNN, and two large-scale datasets, STRINGdb and CPDB.

\end{itemize}

\end{document}